\documentclass[fleqn,usenatbib]{mnras}
\usepackage{multirow}
\usepackage{newtxtext,newtxmath}
\usepackage[section]{placeins}
\usepackage[T1]{fontenc}
\DeclareRobustCommand{\VAN}[3]{#2}
\let\VANthebibliography\thebibliography
\def\thebibliography{\DeclareRobustCommand{\VAN}[3]{##3}\VANthebibliography}
\usepackage{mathtools}
\usepackage{subcaption,graphicx}
\usepackage[normalem]{ulem}
\graphicspath{{img/}}
\usepackage{amsmath}
\usepackage{comment}

\author[Uzeirbegovic et al.]{Emir
  Uzeirbegovic$^{1,2}$\thanks{e.uzeirbegovic@herts.ac.uk}, Garreth Martin$^{3,4}$ and 
  Sugata Kaviraj$^{1,2}$\\
$^{1}$Centre for Astrophysics Research, School of Physics, Engineering \&
Computer Science, University of Hertfordshire, Hatfield, AL10 9AB\\
$^{2}$Centre of Data Innovation Research, School of Physics, Engineering \&
Computer Science, University of Hertfordshire, Hatfield, AL10 9AB\\
$^{3}$Korea Astronomy and Space Science Institute, 776 Daedeokdae-ro, Yuseong-gu, Daejeon 34055, Korea,\\
$^{4}$Steward Observatory, University of Arizona, 933 N. Cherry Ave, Tucson, AZ, USA,\\
}
\pubyear{2021}
\title[Can colours constrain morphology?]{
How the spectral energy distribution and galaxy morphology constrain each other, with application to morphological selection using galaxy colours
}

\begin{document}

\maketitle

\begin{abstract}
We introduce an empirical methodology to study how the spectral energy distribution (SED) and galaxy morphology constrain each other and implement this on $\sim$8000 galaxies from the HST CANDELS survey in the GOODS-South field. We show that the SED does constrain morphology and present a method which quantifies the strength of the link between these two quantities. Two galaxies with very similar SEDs are around 3 times more likely to also be morphologically similar, with SED constraining morphology most strongly for relatively massive red ellipticals. We apply our methodology to explore likely upper bounds on the efficacy of morphological selection using colour. We show that, under reasonable assumptions, colour selection is relatively ineffective at separating homogeneous morphologies. Even with the use of up to six colours for morphological selection, the average purity in the resultant morphological classes is only around 60 per cent. While the results can be improved by using the whole SED, the gains are not significant, with purity values remaining around 70 per cent or below.
\end{abstract}


\section{Introduction}\label{intro}

The visual appearance of galaxies -- commonly referred to as 
galaxy morphology -- correlates with their physical properties,
such as stellar mass \cite[e.g.][]{bundy2005mass}, star formation
rate \citep[SFR, e.g.][]{bluck2014bulge, Kaviraj2014, smethurst2015galaxy, Lofthouse2017},
surface brightness \citep[e.g.][]{martin2019formation, jackson2021origin},
rest frame colour \citep[e.g.][]{strateva2001color, skibba2009galaxy} 
and local environment \citep[e.g.][]{dressler1997evolution, postman2005morphology}.
It reveals key information about the processes that have shaped
the evolution of galaxies over cosmic time \citep[e.g.][]{martin2018role, jackson2020extremely}.
The literature documents many approaches to measuring galaxy
morphology, with historically the most popular being based on 
visual classification schemes \citep[e.g.][]{Hubble1926,Lintott2011,Simmons2017,Kaviraj2019}, 
light distribution based parametric methods \citep[e.g.][]{vaucouleurs1948,Sersic1963,Simard2002,Odewahn2002,Lackner2012,ryan2012size},
and non-parametric approaches such as ‘CAS’ \citep[][]{Abraham1994,conselice2003,Menanteau2006,Mager2018} or
Gini-M20 \citep[e.g.][]{lotz2004,Scarlata2007,Peth2016}. More
recently improving computing power has introduced
new empirical machine learning methods \citep[e.g.][]{Huertas-Company2015,Ostrovski2017,Schawinski2017,Hocking2018,Goulding2018,Cheng2019,Martin2020} often applied to problems of classification or
automated clustering in an attempt to make large surveys
tractable for morphological analysis.

Although estimating the physical parameters of a galaxy
from its spectral energy distribution (SED) is subject to confounders
and ambiguity \citep[e.g.][]{conroy2013, magris2015}, the SED does
trace the stellar composition of the galaxy and relates to physical quantities
like stellar mass, metallicity, ionized-gas properties and star
formation history \citep[e.g.][]{adams2004introduction}. Several
studies exist on the subject of the spectral classification of 
galaxies, which show that spectral types often have corresponding
morphological biases 
\citep[e.g.][]{morgan1957spectral, connolly1994spectral, sodre1994spectral, madgwick2003deep2}. In the same vein, derivatives of the SED such as colours, are thought
to correlate strongly with morphology and have frequently
been used by astronomers as a proxy thereof (see \cite{masters2019galaxy}
for an extensive list). However, given that the shape of a SED
is driven by the galaxy's star formation history, whilst morphology
depends on dynamical factors, it is not clear to what extent the
SED and morphology can be expected to constrain each other.

Results from the spectral classification literature show that
morphologies associated to spectral types are impure to varying
degrees. For example, \cite{masters2010galaxy} morphologically
select 5433 face on spirals from the SDSS Data Release 6
\citep{adelman2007fifth} of which 6 per cent are red spirals,
which runs contrary to the expectation that late-types
should have bluer colours than early-types. They also show
that red spirals primarily manifest themselves in the higher
mass range, and motivate the claim that massive galaxies are
red regardless of morphology. \cite{bamford2009galaxy} show that
colour and morphology appear to be differently related to the
local environment and are bound to stray from a one-to-one 
mapping. Although these results imply that the constraints
of SED and morphology on each other are asymmetrical, the 
question is yet to be explored directly and the constraint of
morphology on the SED is yet to be properly considered.

In this work we introduce an empirical methodology for the 
analysis of how SEDs and morphologies constrain each other.
We utilise the vector space building techniques for galaxy
surveys laid out in \cite{uzeirbegovic2020}. Galaxy cutouts
are projected to points in a high dimensional space in which
the Euclidean distances between galaxies relate directly 
to the visual morphological similarity between them. We 
likewise project galaxy SEDs as proxied by the catalogued 
rest-frame bands. We use these vector spaces together to
demonstrate the implications that the structure of one space
has on the other. 
In contrast to classification or morphological indicator-based methods, vector spaces enable us to compare galaxy 
similarity on a continuous scale, and -- as we will show
-- conduct analyses without having to commit to specific 
features such as CAS or a particular typology
 \citep[e.g. the Hubble sequence;][]{Hubble1936}.

From here, the paper is structured as follows. In 
Section \ref{prep}, we describe the source data used, how it
is prepared, how the key artefact -- the dissimilarity 
matrices -- are derived from it, and how they are interpreted.
In Section \ref{analysis} we analyse the vector spaces to 
show that: (1) SEDs significantly constrain morphology and
the extent to which it happens can be quantified, (2) the
constraint is not uniform and some morphologies are 
especially well-constrained by the SED, (3) galaxy morphology
constrains SEDs more than vice versa, (4) there is a 
visually distinguishable group of galaxies for which the
SED and morphology are mutually constraining, and (5) there
is an upper bound on coverage and purity when attempting
to use rest-frame colours to select homogeneous morphologies, with direct implication on colour based morphology selection
methods. Finally, in Section \ref{conclusion} we conclude by
summarising the main points of our work. 

\section{Data and preparation}\label{prep}

\subsection{HST-CANDELS}

We use the {\it HST} CANDELS \citep{grogin2011candels,koekemoer2011candels}
survey because it offers a high-resolution probe of galaxy
evolution. It is supplied with a corresponding catalogue
\citep[]{guo2013, santini2015}. The survey consists of 
optical and near-infrared (UVIS/IR) images from the Wide
Field Camera 3 (WFC3) and optical images from the Advanced
Camera for Surveys (ACS) in five well-studied extragalactic
survey fields. We focus on GOODS-S, one of the deep tier
(at least four-orbit effective depth) fields. In order to
keep galaxies comparable to each other, and minimise the effects
of noise, we select only galaxies
imaged in the WFC3 F160W filter at $z<3$ with signal-to-noise
ratio $>20$ and $M_{\star}\ge 10^8$ $\rm{M_{\odot}}$.
The signal-to-noise ratio is provided per object in
the catalogue and is calculated as flux divided by flux error
in the F160W band. This filtering results in a sample of
7757 galaxies with images in the WFC3 F160W filter.

For each object after filtering, we take $31\times 31$
(1.8$^{\prime\prime}$) pixel cutouts, using the catalogued
sky coordinate as a centroid. We also make use of the
photometric redshift, mass, star-formation rate and
rest-frame $UBVRIJK$ magnitudes provided in the catalogue. The photometric redshifts have NMAD values better than 0.05 and 0.03 at $z<1.5$ and $z>1.5$ respectively and outlier fractions of $\sim$5 per cent or better. From herein, we refer to the total number of galaxies in our dataset as $N$, where $N=7757$, and the pixel width of square cutouts as $n$, where $n=31$.

\subsection{The morphological dissimilarity matrix}\label{viz-matrix}

The representation of (visual) morphological similarity in our method is
based on the calculation of a ``dissimilarity'' matrix, which
encodes a measure of the pairwise differences between galaxies.
To create a dissimilarity matrix we begin by projecting all 
galaxy cutouts onto a common vector space, in which the squared
Euclidean distance between vectors is informative of morphological
differences. We call this the \textit{morphology space}. We follow
the vector space building methodology for galaxy surveys laid out
in \cite{uzeirbegovic2020} with the following cutout 
standardisation steps applied sequentially prior to projection:
\begin{enumerate}
\item Masking - Many cutouts contain background noise which can
  cause spurious similarity. For each cutout, we mask away the
  background by robustly fitting a Gaussian using the median
  ($\alpha$) and interquartile range ($\beta$) of the flux
  densities of the pixels in the cutout, and clipping to zero
  all pixels below a threshold $t$ given by the solution to
  $\Phi(X\le t\mid \alpha,\beta)=0.9$, where $\Phi$ is a Gaussian
  CDF. Since a majority of the pixels in most cutouts are 
  background dominated, this has the effect of zeroing out the
  background. The process is not overly sensitive to our chosen
  constant ($0.9$) and will produce much the same results in the
  range $0.85-0.95$.
\item Rotation - The covariance of the coordinates of all the
  non-zero pixels in F160W are used to find the major axis of the
  data. The cutout is then rotated so as to bring the galaxy in 
  line with the major axis. This step standardises the orientation
  of all cutouts.
\item Flipping - All cutouts are flipped as necessary horizontally
  and vertically to make sure that the brightest pixels are in
  the top left hand corner. This makes it more likely that bright
  spots in similar galaxies line up. This is achieved by comparing
  the top/bottom, left/right sides of each cutout respectively and
  flipping accordingly to move the brightest quadrant to the top
  left of each cutout.
\item Normalisation - The scale of flux densities in individual images is removed by
  normalising each cutout to the range $[0,1]$.
\end{enumerate}
We project the $N$ standardised cutouts by flattening each of
them into row vectors and then stacking the vectors into a
$N \times n^2$ matrix. We use Principal Component Analysis (PCA)
to decompose the matrix into orthogonal basis vectors. We test
that PCA is not degenerate by repeatedly leaving out 20 per cent
of galaxies at random and making sure the fitting does not change
substantially as described in \cite{uzeirbegovic2020}. We decide
to retain $k=40$ basis vectors by trying each value of $k$ in
turn and picking the point at which additional dimensions
stop making a difference to how the dissimilarities are 
distributed.

A dissimilarity of zero implies that two galaxies are the same,
and otherwise the closer to zero the more similar galaxies are.
That is enough for our analysis as we are only interested in the
rank of a galaxy relative to another (detailed in Section
\ref{analysis}), and not the absolute distances. However, for
the astronomer in search of a more intuitive interpretation of
dissimilarities it should be noted that
$Cor(X,Y) \propto 1-d^2(\bar{X},\bar{Y})$ where $X,Y$ are row
vectors, $\bar{X},\bar{Y}$ are z-scaled versions, and $d^2$ is
the squared Euclidean distance function. That is, a dissimilarity 
may be interpreted as the correlation between the pixels in
two images.

\subsection{The SED dissimilarity matrix}\label{sed-matrix}

We use rest-frame magnitudes in the $UBVRIJK$ bands as a proxy
for the SED of any given galaxy. We shall refer to this proxy
as the SED for brevity. One objection may be that a 7D
proxy of the SED underestimates the variance available in a
fuller sampling of the spectrum. To investigate, we conduct a
principal component analysis of the UBVRIJK bands. The first
component accounts for over 96 per cent of explained variance,
increasing to 99.9 per cent when a second component is added.
The extreme redundancy suggests there would be limited benefit
to a finer sampling of the spectrum between bands $U$ and $K$.
However, if the spectrum beyond $U-K$ is discontinuous with
$U-K$, then it could affect the results.

The SEDs are treated as 7D vectors and the SED dissimilarity
matrix is calculated as the pairwise squared Euclidean distances
between all SEDs. We call this the \textit{SED space}. Note 
further that since colour is just the magnitude in two bands subtracted from each other 
(e.g. $U-R$), nearby SEDs imply similar bands which in turn
imply similar colours on average. It may thus help readers to
think about SED distance as an indicator of how closely matched
two galaxies are across all the colours implied by the rest-frame
bands in use.

\section{Analysis}\label{analysis}

In this section we show how SED and morphological dissimilarity
matrices can be used together to produce insights regarding
the relationship between the SED and morphology. We
note that all the graphs featuring density plots are produced by
resampling the original series 1000 times, calculating 1000
histograms and reporting back the 16th-84th percentiles as an
area to give the reader a clearer sense of sample uncertainty.
The densities per histogram are just the underlying frequencies, normalised such that the integral over all the bins equals 1.

\subsection{SED constraints on morphology}

To determine whether a galaxy's SED constrains its morphology,
it helps to consider the extreme cases wherein (1) the SED
completely determines the morphology, and (2) the SED is 
completely independent of morphology. Suppose we draw a galaxy
at random and then find its nearest neighbour in SED space
(that is, another galaxy which has a SED most similar to its
own). In case 1, we may expect the two galaxies to look very
similar (that is, to have a small morphological dissimilarity
distance), whilst in case 2 we may expect that the morphological
dissimilarity distance could be anything at all.

We generalise this intuition as follows. For every galaxy $g$,
we find its nearest neighbour in SED space, $\bar{g}$. We then
order all the galaxies in morphology space by distance to $g$
and calculate the fraction $x$ that are closer to $g$ than
$\bar{g}$ is. For example, if for some galaxy $g$, the
nearest neighbour in SED space $\bar{g}$ has $x=0.1$, it means
that 10 per cent of other galaxies are more visually similar to
$g$ than $\bar{g}$ is. We call $x$ the \textit{morphology space
  fraction}. It should be noted that all comparisons and
nearest neighbour searches are restricted to be within $0.15$
of the photo-z of $g$ to keep comparisons within similar visual
conditions. In case 1 we should expect $x=0$. In case 2 -- since
$\bar{g}$ might end up ranked anywhere in the ordered set -- we
should expect $x=0.5$ on average. We calculate $x$ as above, for
every galaxy. 
\begin{figure}
  \centering
  \includegraphics[width=8cm]{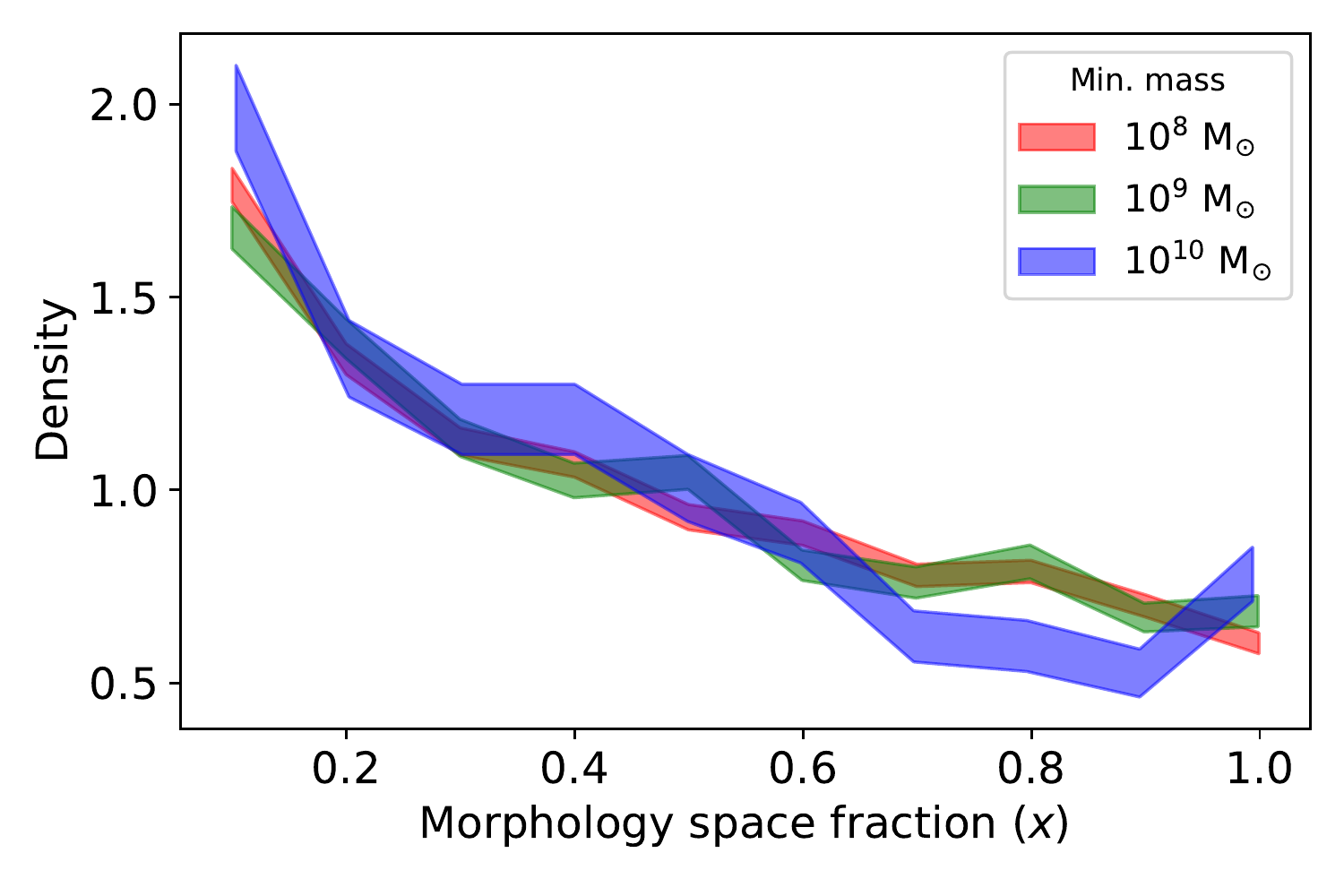}
  \caption {
    A density plot of the distribution of the morphology
    space fraction for various stellar mass cuts. The areas
    represent the density between the 16th and 84th percentiles.
    All distributions are heavily skewed towards smaller
    fractions, peaking at $x<0.1$. The skew can be seen to
    increase at the highest mass cut, which may be because larger
    galaxies are better resolved and therefore less ambiguous
    in morphology space. The overall median $x$ value is $0.37$,
    and the smallest sample size is 1222. The results are
    significant at a p-value approaching zero.
  }
  \label{frac-dist}
\end{figure}
Figure \ref{frac-dist} shows a density plot of the resultant
distribution at various mass cuts. All distributions are 
heavily skewed towards smaller fractions, peaking at $x<0.1$.
The skew can be seen to increase at the highest mass cut,
which may be because larger galaxies are better resolved and
therefore less ambiguous in morphology space. The overall 
median $x$ value is $0.37$ and the smallest sample size is 
1222. For any given mass cut, it is straightforward to show
using the binomial distribution with $p=\frac{1}{m}$, where
$m$ is sample size, that the probability of these results
occurring by chance is approximately zero. Thus, SED
similarity between galaxies implies a significantly greater
likelihood of similar morphology.

Another useful way to view the result is in terms
of the probability that galaxies which are most similar to
each other via the SED are also most similar to each other via
visual morphology. Let $\epsilon$ be a morphology space
fraction threshold below which a galaxy is considered
suitably similar. For example, $\epsilon \le 0.01$ implies
that there are 1 per cent or fewer galaxies more similar to
$g$ than $\bar{g}$ is. Under the null hypothesis that the
SED and morphology are independent, the probability that
$\bar{g}$ (the most SED similar galaxy to $g$) is visually
similar enough, is $\epsilon$. In our data, at 
$\epsilon = 0.005$, the empirical probability is 
approximately $3\epsilon$: an improvement by a factor of 3.

In the next section we consider the group of galaxies at $x < 0.1$
wherein the SED is particularly constraining.

\subsection{The $x<0.1$ group in morphology space}

In this section we consider in more detail the galaxies for
which morphology appears particularly well constrained by the
SED.
\begin{figure}
  \centering
  \includegraphics[width=8cm]{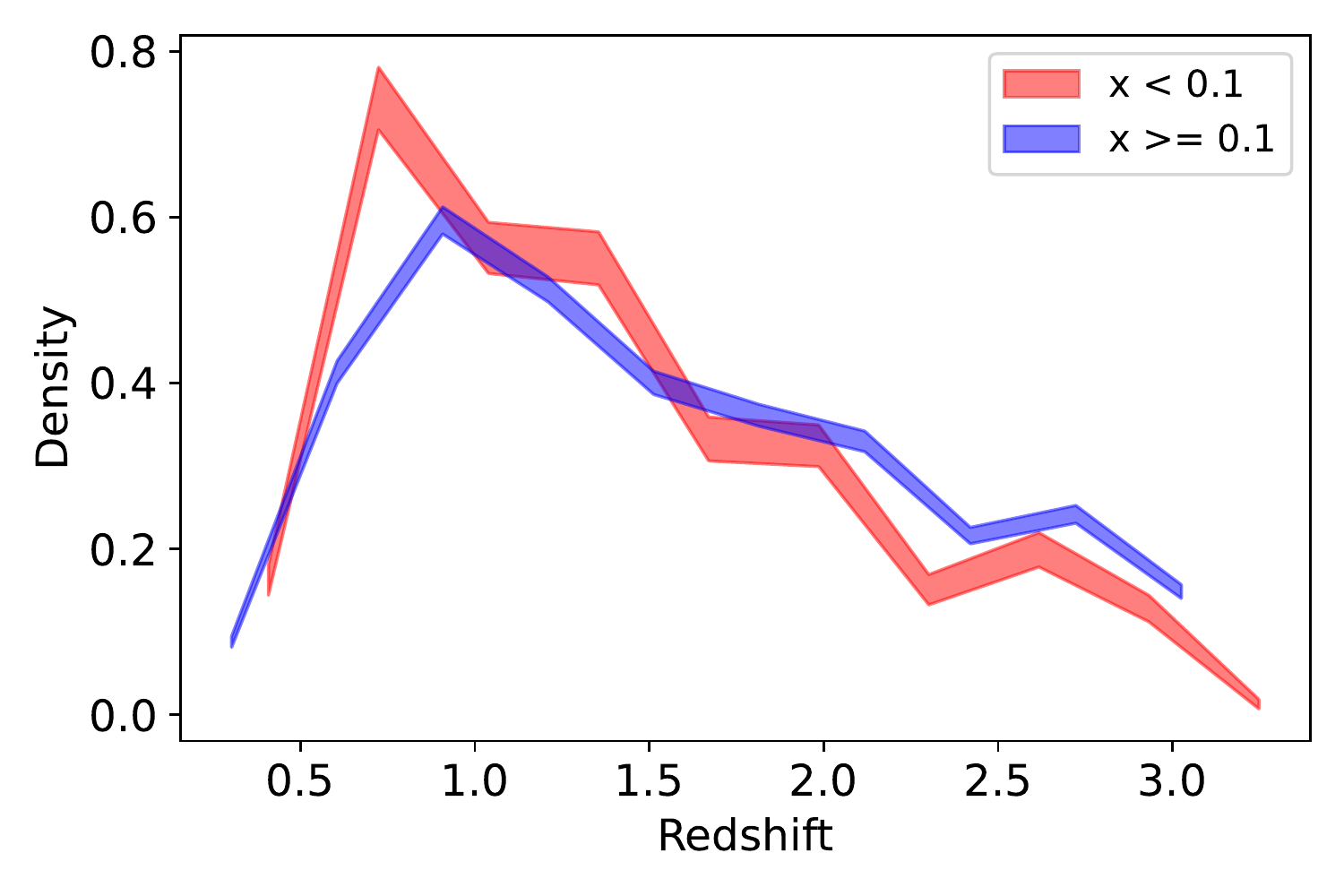}
  \includegraphics[width=8cm]{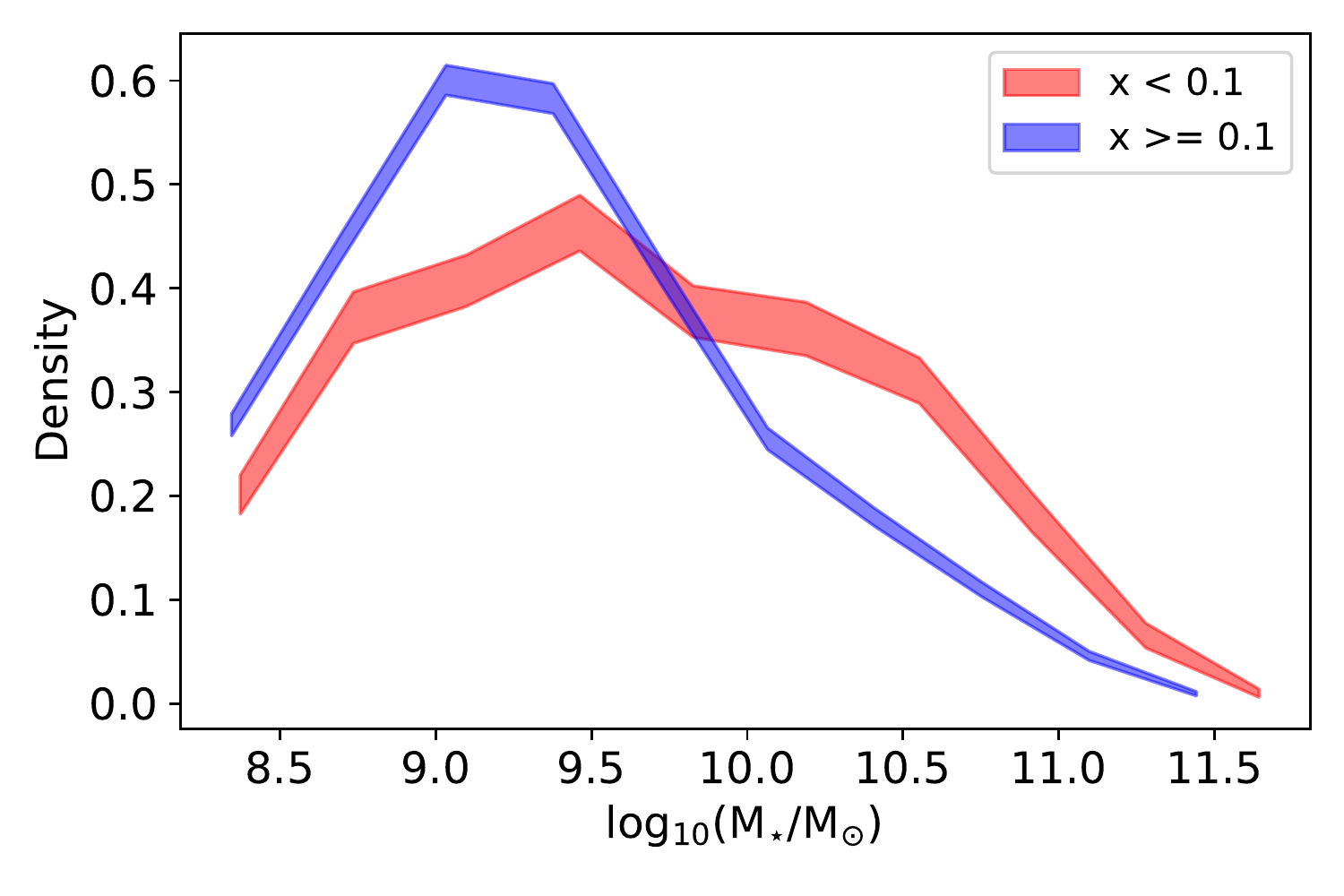}
  \includegraphics[width=8cm]{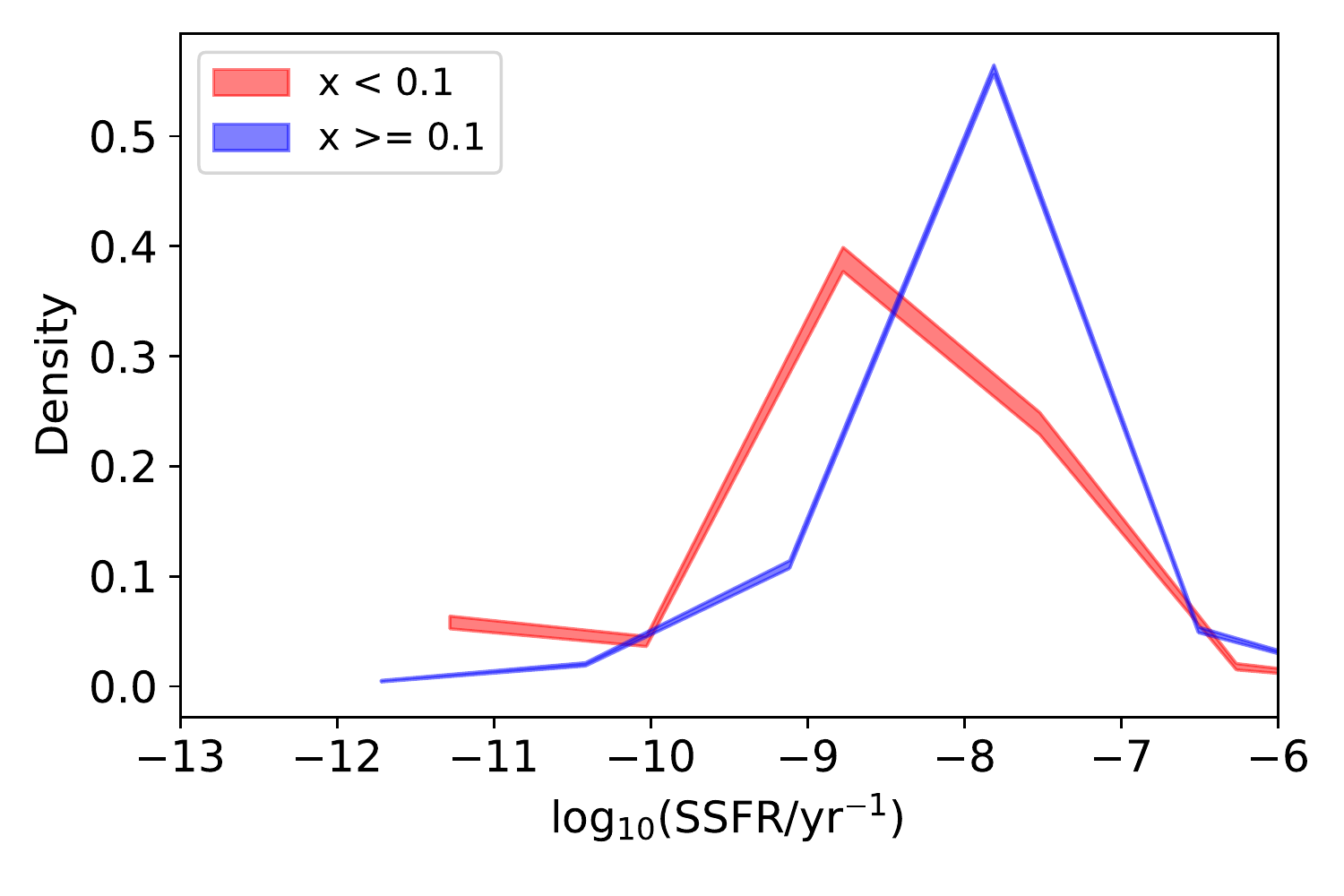}
  \includegraphics[width=8cm]{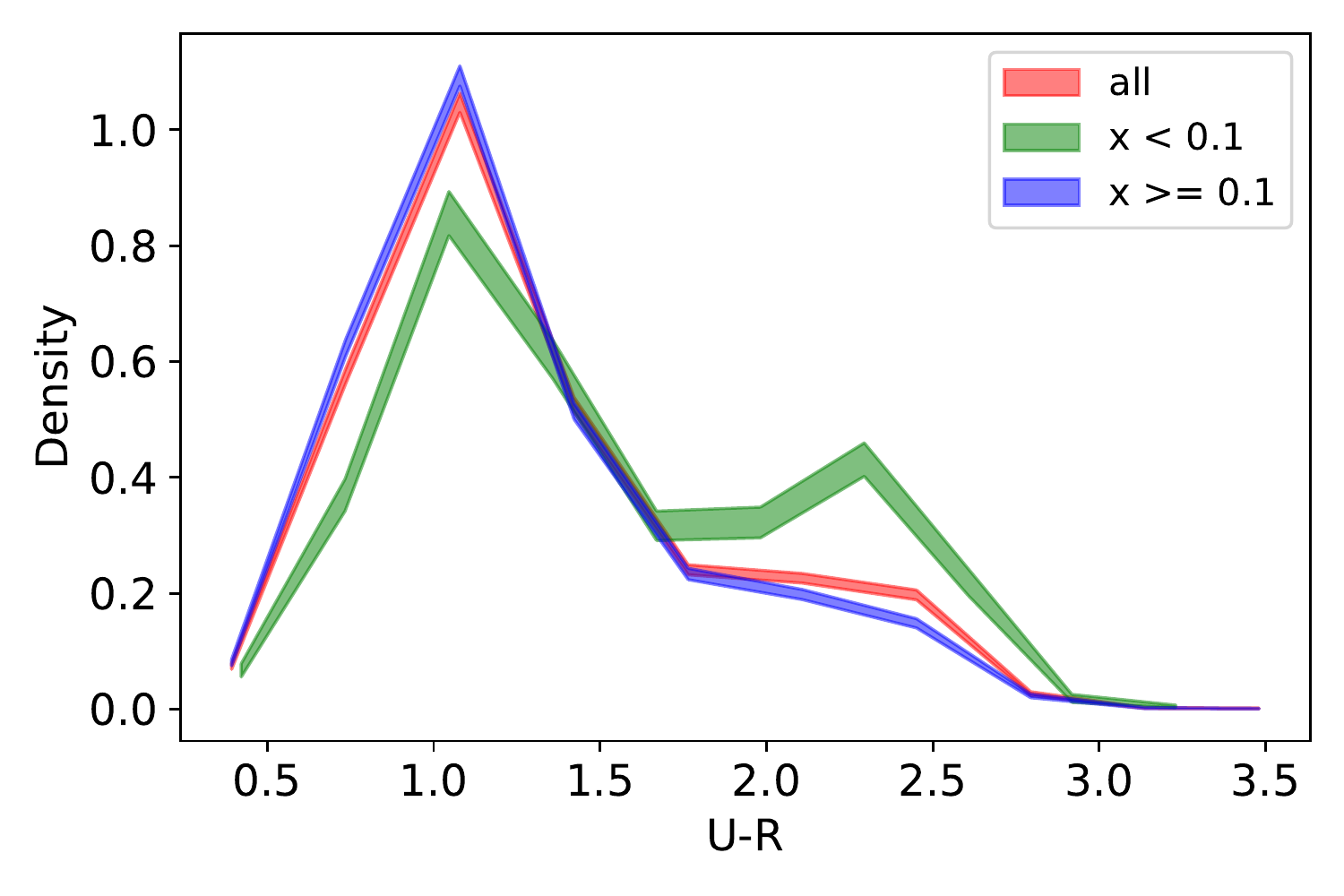}
  \caption {
    Comparisons of redshift, stellar mass, SSFR and rest-frame
    $U-R$ colour for different groups. The areas represent the
    density between the 16th and 84th percentile.The $x<0.1$
    group is over represented in the redshift range $0.25 < z < 1$ and at higher
    stellar masses but present throughout the redshift and mass range
    of the population. It has a lower SSFR and a significantly
    different $U-R$ colour distribution wherein bimodality 
    with peaks at 1 and 2.1 is clearly present for the 
    $x<0.1$ group, in contrast to the unimodal distribution of the $x\ge0.1$ group.
  }
  \label{group-comp}
\end{figure}
Figure \ref{group-comp} shows group comparisons of redshift,
mass, specific star formation rate (SSFR) and $U-R$ colour.
The $x<0.1$ group is over-represented in $0.25 < z < 1$ and at
higher masses but present throughout the redshift and mass
range of the population, which indicates that this is not an
artificial result brought about by the fact that more nearby
objects are better resolved and therefore easier to 
differentiate. This group has a lower SSFR and a significantly
different rest-frame $U-R$ colour distribution which is
bimodal, compared
with the unimodal $x\ge 0.1$ distribution. There are 1390 
galaxies in the $x<0.1$ group, which comprises around 18
per cent of the dataset. Thus, the galaxies for which morphology
has the greatest link with the SED are typically
redder and more massive compared with more visually dissimilar
objects.

We next consider how this group differs from the rest of the
galaxies in terms of morphology. In order to investigate
possible patterns in morphology amongst the $x<0.1$ group,
we empirically generate a set of morphological classes and
then cross tabulate these against $x<0.1$ group membership
to see whether in-group galaxies are over/under represented
anywhere. 

We first explain how the morphological classes are generated
and then how they are used in our analysis. We use the
morphology space to naturally partition galaxies around a set
of exemplars (i.e. prototypical galaxies) such that galaxies in the
same partition are all close to the same exemplar. Let
$\{d_{i,j}\}$ be our $N \times N$ dissimilarity matrix in
morphology space, the objective then is to choose a set of
exemplars such that the sum of the distances between each
galaxy and its closest exemplar is minimised. Formally,
for some set of galaxy vectors $s\in S$:
\begin{equation} \label{apeq}
  \min_{\{q_i\}_{i=1}^m}\Bigg( \sum_{s\in S} \min_i{d_{s,q_i}} + \sum
  d_{q_i,q_i}\Bigg)
\end{equation}
where $q_i,...,q_m\in S$ are $m=|S|$ examplars. The diagonal
$d_{q_i,q_i}$ is set to the median distance. The optimisation
above is usually intractable but the Affinity Propagation
\citep[AP;][see Appendix \ref{affinity} for a short
overview]{dueck2009affinity} algorithm offers a good 
approximation. A crucial advantage with AP is that the number
of partitions are automatically determined.
\begin{figure}
  \centering
  \includegraphics[width=8cm]{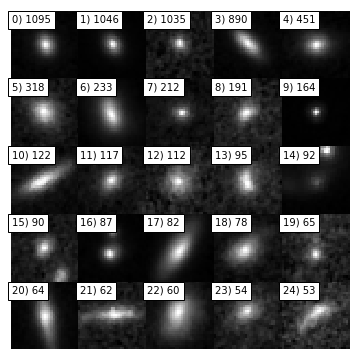}
  \caption {
    The exemplars (prototypical galaxies) of the top 25 
    partitions ordered by size from left to right, top to
    bottom. The labels indicate the partition index and the
    number of galaxies in the partition. All partitions
    contain more than 50 members and together make up more
    than 90 per cent of the dataset.
  }
  \label{exemplars}
\end{figure}
We run AP on our dataset which returns 68 partitions, of which
25 have at least 50 galaxies and subsume more than 90 per cent
of the dataset.
Figure \ref{exemplars} shows the top 25 partitions along with
their indices and the number of galaxies in them. Full panels
containing all the galaxies in each partition are available as
part of the supplementary material.
\begin{table}
  \centering
  
\begin{tabular}{rrrrrrr}
\hline
   Part. &   $x\ge 0.1$ &   $x<0.1$ &   \# &    z &    mass &   $(U-R)>2$ \\
\hline
       0 &    0.86 &  -1.83 & 171 & 1.02 & 1.64e+09  &      40 \\
       1 &   -1.3  &   2.78 & 226 & 1.16 & 7.58e+09 &     103 \\
       2 &    0.72 &  -1.53 & 165 & 1.41 & 1.04e+09  &      35 \\
       3 &    0.73 &  -1.57 & 140 & 0.93 & 1.70e+09 &       9 \\
       4 &    0.52 &  -1.11 &  71 & 1.03 & 6.64e+09  &      21 \\
       5 &   -0.06 &   0.12 &  58 & 0.86 & 5.07e+09 &       3 \\
       6 &   -0.37 &   0.8  &  47 & 0.93 & 4.68e+09  &       5 \\
       7 &    0.69 &  -1.47 &  29 & 1.47 & 1.33e+09  &       7 \\
       8 &    0.74 &  -1.59 &  25 & 1.57 & 9.74e+08  &       1 \\
       9 &   -2.55 &   5.45 &  59 & 0.5  & 1.28e+09  &      31 \\
      10 &   -2.11 &   4.51 &  43 & 0.87 & 6.82e+09  &       4 \\
      11 &    0.21 &  -0.44 &  19 & 1.41 & 1.22e+09  &       1 \\
      12 &   -0.2  &   0.42 &  22 & 1.09 & 4.44e+09 &       2 \\
      13 &    0.35 &  -0.74 &  14 & 1.51 & 1.33e+09  &       0 \\
      14 &   -0.29 &   0.61 &  19 & 2.03 & 3.89e+08  &       2 \\
      15 &    0.02 &  -0.04 &  16 & 0.95 & 4.17e+08 &       0 \\
      16 &    0.9  &  -1.93 &   8 & 1.97 & 4.46e+09  &       1 \\
      17 &   -0.52 &   1.11 &  19 & 0.74 & 6.64e+09  &       2 \\
      18 &    0    &  -0    &  14 & 1.03 & 2.67e+09 &       1 \\
      19 &    0.91 &  -1.95 &   5 & 1.92 & 3.13e+09  &       0 \\
      20 &   -1.31 &   2.8  &  21 & 0.63 & 6.96e+09  &       4 \\
      21 &    0.16 &  -0.34 &  10 & 1.52 & 2.12e+09 &       0 \\
      22 &   -0.32 &   0.68 &  13 & 0.68 & 3.7e+09   &       0 \\
      23 &   -0.05 &   0.1  &  10 & 1.2  & 1.40e+10 &       0 \\
      24 &    0.99 &  -2.11 &   3 & 1.02 & 1.09e+09  &       1 \\
\hline
\end{tabular}
\caption {
  The table shows the top 25 partitions by size along with
  the Pearson's residuals for the $x\ge0.1$ and $x<0.1$ groups
  where $x$ is the morphology space fraction. Additionally, 
  for the $x<0.1$ group, the table shows the sample size, mean
  redshift, mean stellar mass and the number of galaxies with
  a $U-R$ > 2. Its noteworthy that the $x<0.1$ group is
  significantly over/under represented in several partitions, 
  and that the redder peak in the colour distribution given in
  Figure \ref{group-comp} is mostly accounted for by the
  over-representation in partition 1.
}
\label{prominence}
\end{table}

We next produce a cross tabulation of all the galaxies,
counting each into a partition and either in or out of 
group $x<0.1$. We then conduct a contingency table 
analysis, producing Pearson's residuals for each cell. A
Pearson residual can be interpreted as the difference
(measured in standard deviations) between the observed and
expected counts in each cell, on the assumption (for the
expected counts) that there is no dependence between the
partitions and the $x<0.1$, $x\ge0.1$ groups. For example,
a residual of 2 or more could be interpreted as a likely
over-representation of galaxies in that cell, and a residual
of -2 or less could be interpret as an under representation. 

Table \ref{prominence} shows the top 25 partitions by size
along with the Pearson's residuals for the $x\ge0.1$ and
$x<0.1$ groups. Additionally, for the $x<0.1$ group, the
table shows the sample size, mean redshift, mean stellar
mass and the number of galaxies with rest-frame $U-R>2$.
Its noteworthy that the $x<0.1$ group is significantly
over/under represented in several partitions, and that
the redder peak in the colour distribution from Figure 
\ref{group-comp} is mostly accounted for by the 
over-representation in Partition 1. Thus, SED constrains
morphology most strongly for galaxies that are typically
more massive, redder and morphologically elliptical.

\subsection{Morphology constraints on SED}

We now consider how SEDs may constrain morphology.
\begin{figure}
  \centering
  \includegraphics[width=8cm]{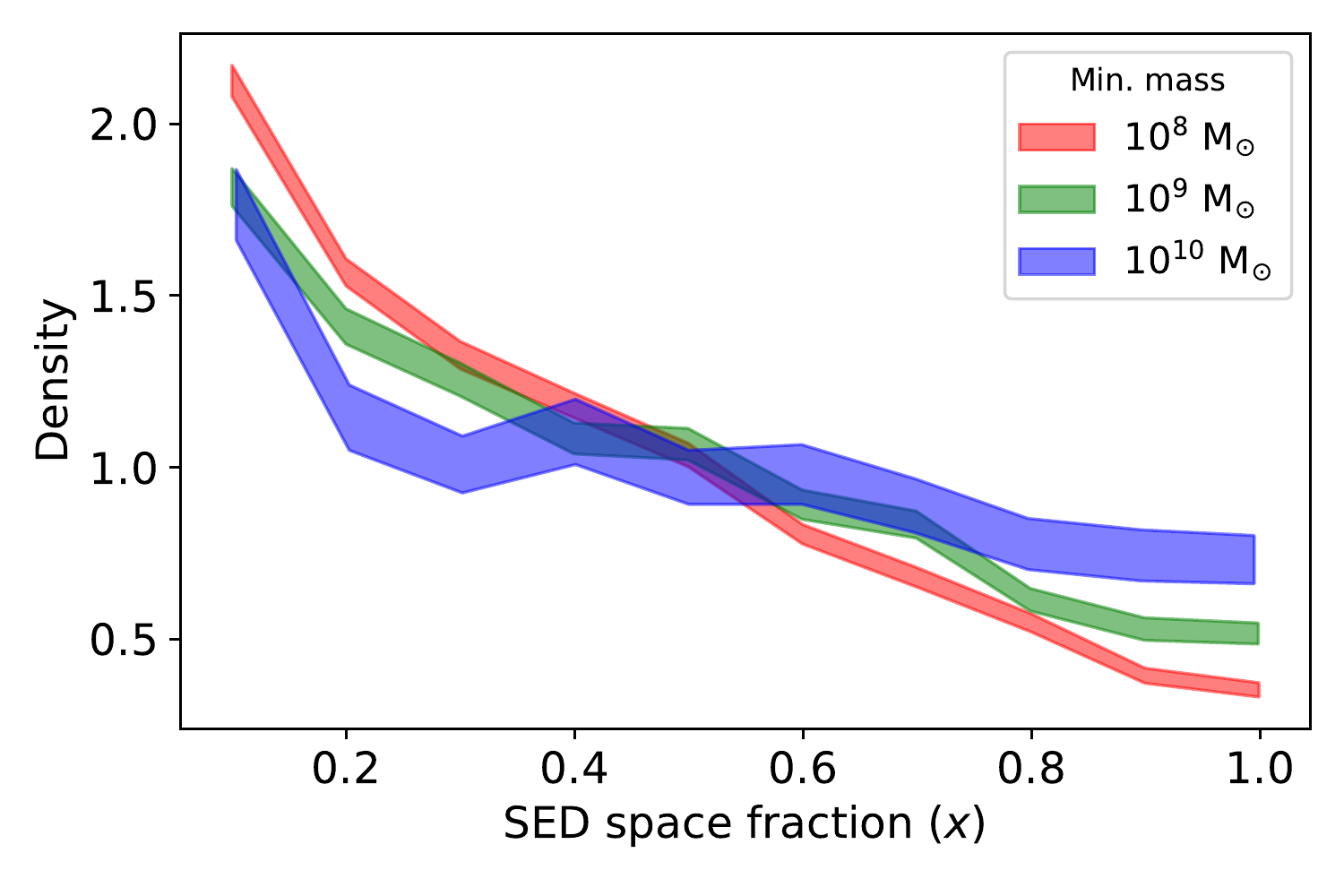}
  \caption {
    The SED space fraction density distribution at a
    selection of mass cuts. The areas represent the density
    between the 16th and 84th percentile. Unlike Figure
    \ref{frac-dist}, as the mass minimum increases, the left
    skew decreases, likely because more massive galaxies have
    less diverse SEDs therefore increasing ambiguity. Overall
    the median $x$ value is $0.30$, and the smallest sample
    size is 1222.
  }
  \label{frac-dist-appear}
\end{figure}
Just as before, but in the opposite direction, for every 
galaxy $g$, we find its nearest neighbour in morphology 
space $\bar{g}$. We then order all the galaxies in SED 
space by distance to $g$ and calculate the fraction $x$ 
that are closer to $g$ than $\bar{g}$ is. We call this 
the \textit{SED space fraction}. It should be noted that
all comparisons and nearest neighbour searches are 
restricted to be within $0.15$ of the photo-z for $g$ to
keep comparisons within similar visual conditions. Figure
\ref{frac-dist-appear} shows the distribution of the SED
fraction at various mass cuts. Unlike Figure \ref{frac-dist},
as the minimum mass increases, the left skew decreases, 
likely because more massive galaxies have less diverse
SEDs therefore increasing ambiguity. It should be noted
that, as with the morphology space fraction distribution,
the SED space fraction distribution is most dense at
$x<0.1$. Overall the median $x$ value is $0.30$, and the smallest sample size is 1222.

As before, another useful way to view the result
is in terms of the probability that galaxies which are
most similar to each other via morphology are also most
similar to each other via SED. Let $\epsilon$ be a SED
space fraction threshold below which a galaxy is considered
suitably similar. Under the null hypothesis that the SED
and morphology are independent, the probability that
$\bar{g}$ (the most morphologically similar galaxy to $g$)
is SED similar enough, is $\epsilon$. In our data, at
$\epsilon = 0.005$, the empirical probability is
approximately $2.75\epsilon$: an improvement by a factor
of 2.75.
\begin{figure}
  \centering
  \includegraphics[width=8cm]{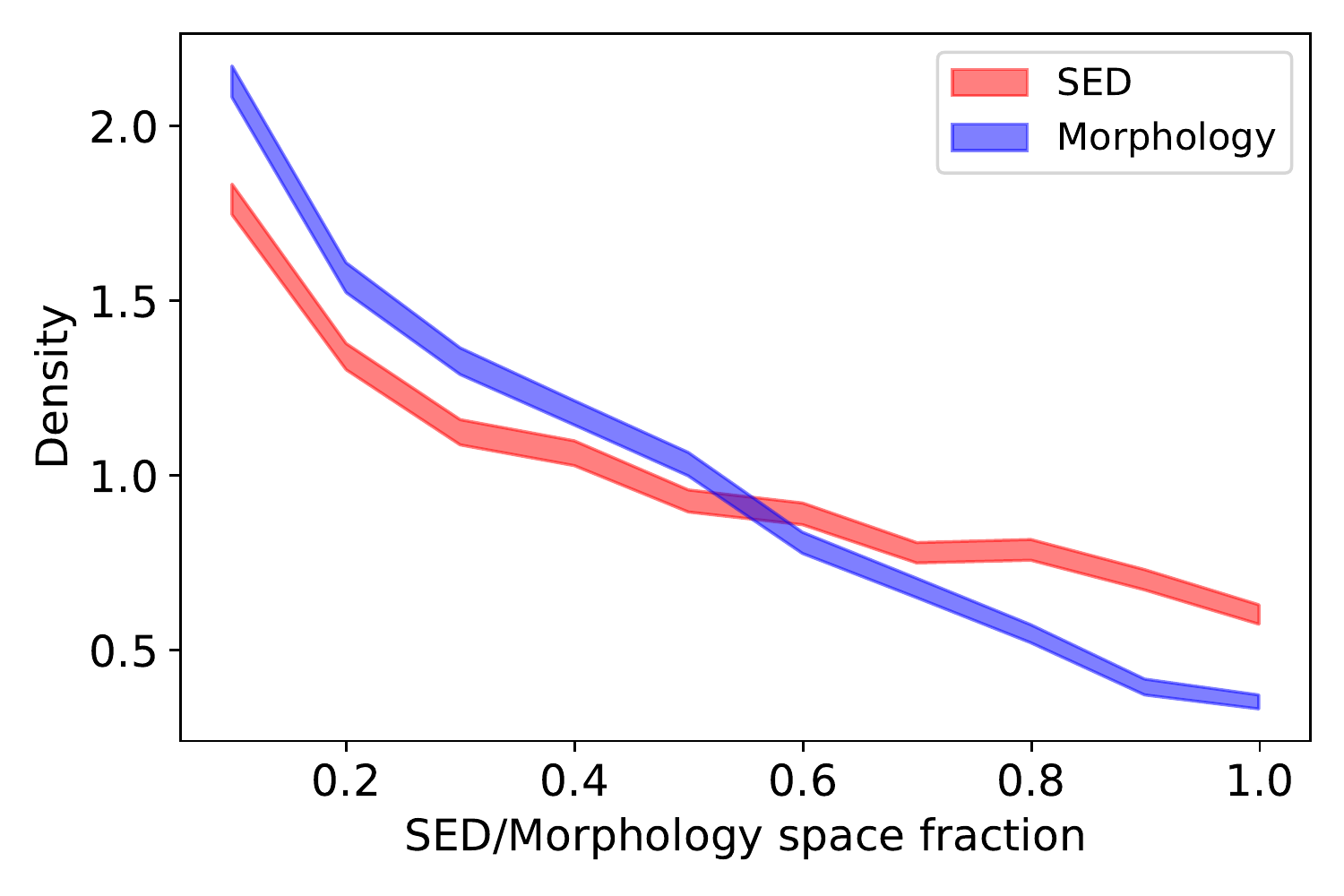}
  \includegraphics[width=8cm]{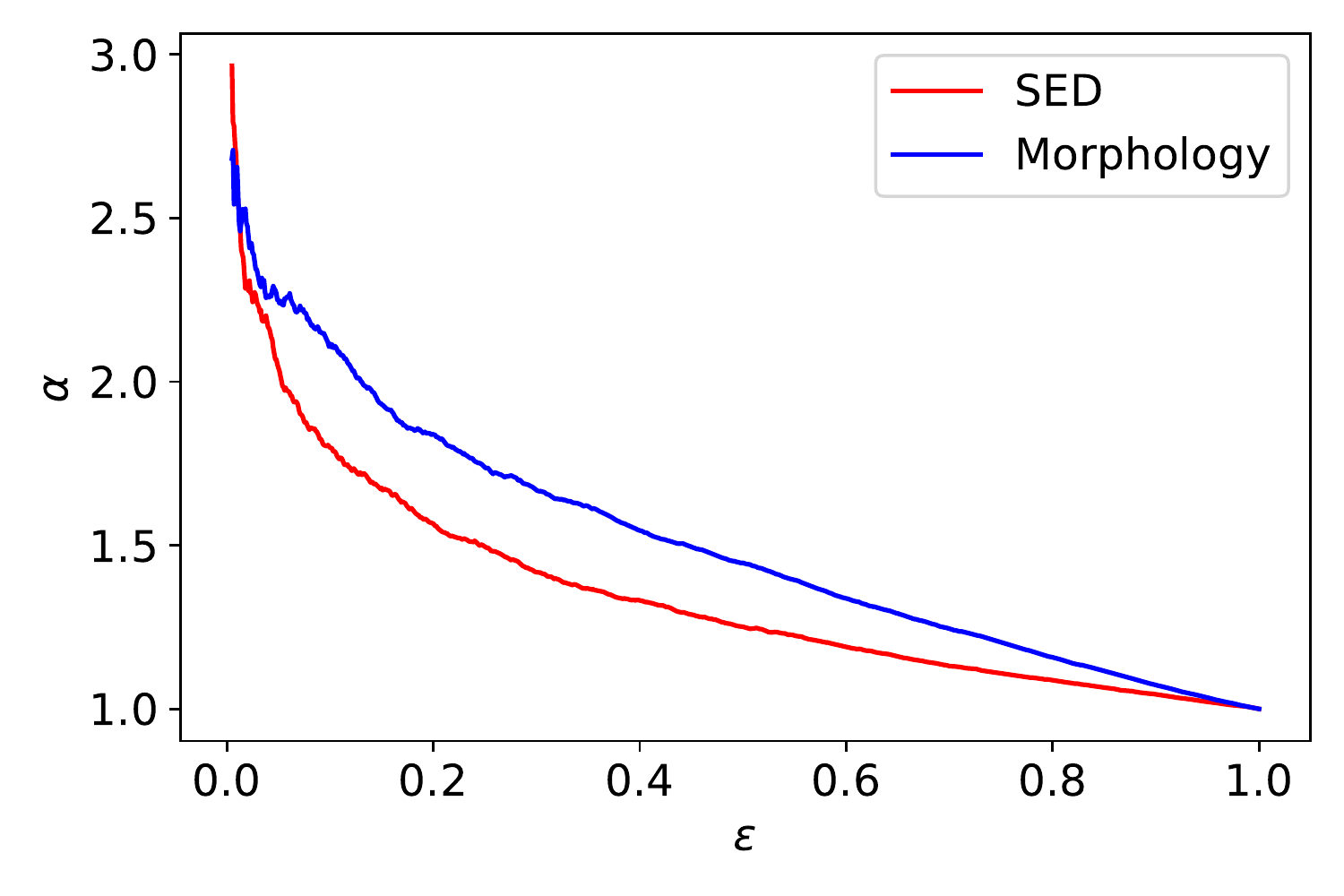}
  \caption {
    Density plot (top) of morphology fraction distribution
    and SED fraction distribution. The areas represent the
    density between the 16th and 84th percentile.Both series
    show a definitive skew towards smaller fractions. However,
    it is noteworthy that on average morphology constrains
    the SED significantly more than vice versa: the median
    fractions are 0.30 for morphology and 0.37 for SED.
    The bottom plot shows the relative gain in probability
    ($\alpha\epsilon)$ of $\bar{g}$ being within the
    $\epsilon$ threshold. It shows that even though the
    SED constraint results in a higher peak probability,
    morphology constraints are more limiting over all but
    the very beginning of the threshold range.
  }
  \label{sed-vs-appear}
\end{figure}

Figure \ref{sed-vs-appear} shows the overall morphology
fraction distribution and SED fraction distribution. Both
density series show a definitive skew towards the smaller
fractions. However, it is noteworthy that morphology
constrains the SED significantly more than vice versa
on average (0.30 versus 0.37 median fractions). The
bottom plot shows the multiple of the gain in probability
compared to the null hypothesis (i.e. ($\alpha\epsilon)$
of $\bar{g}$ being within the $\epsilon$ threshold). It
shows that even though the SED constraint results in a
higher peak probability, morphological constraints are
more limiting over all but the very beginning of
the $\epsilon$ range.

\subsection{A mutually constrained group}

In this section we investigate whether the galaxies
which fall into the $x<0.1$ group in both the morphology
and SED spaces are different from the rest.
\begin{figure}
  \centering
  \includegraphics[width=8cm]{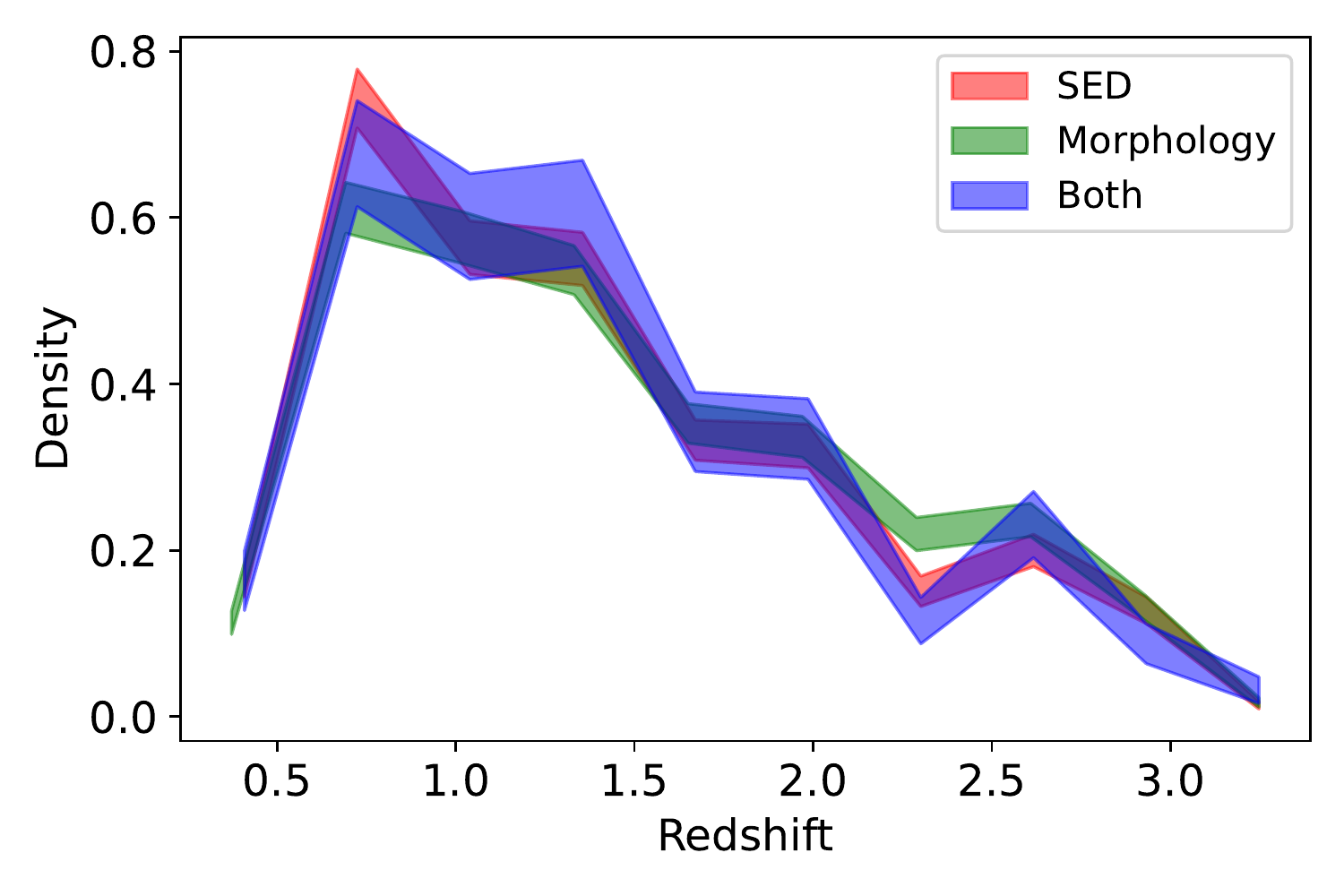}
  \includegraphics[width=8cm]{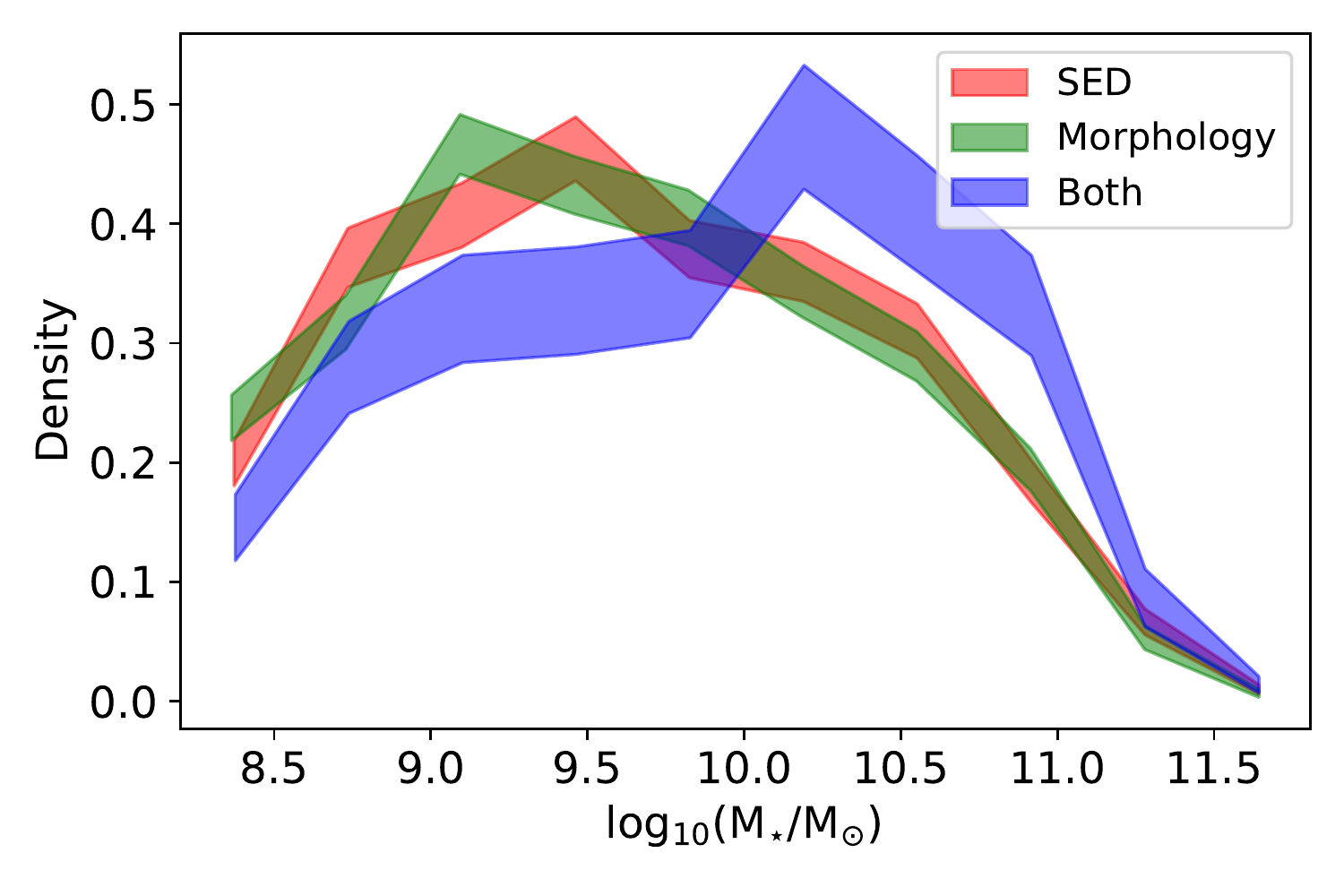}
  \includegraphics[width=8cm]{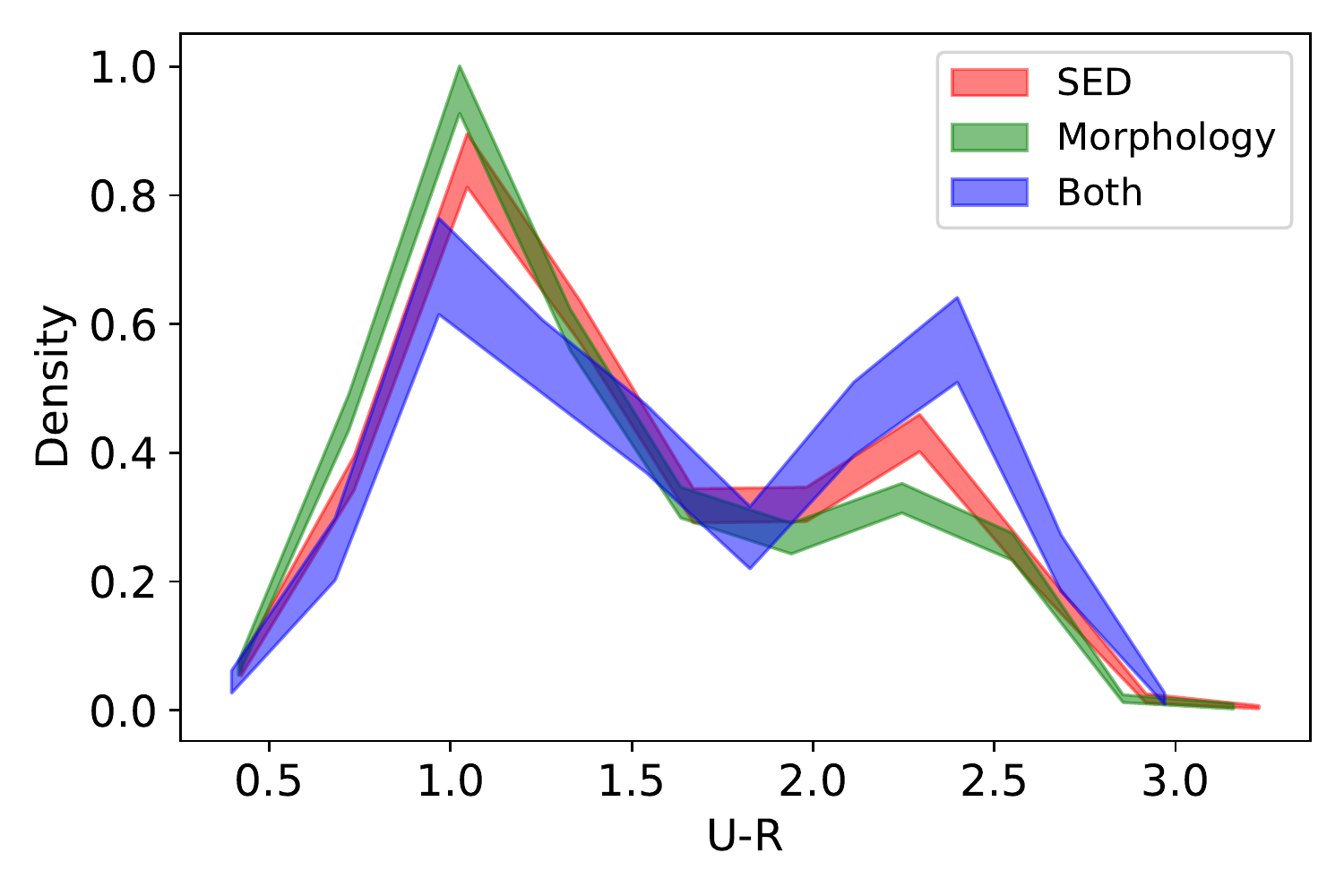}
  \caption {
    The redshift, mass and rest-frame $U-R$ colour of
    galaxies in the $x<0.1$ group by SED and morphology,
    and in both. The areas represent the density between
    the 16th and 84th percentile. Its noteworthy that the
    ``Both'' subset contains galaxies that are more 
    massive and bluer than the other two.
  }
  \label{sed-appear-both}
\end{figure}
Figure \ref{sed-appear-both} shows the redshift, stellar
mass and colour distributions for the $x<0.1$ group in 
morphology space, SED space, and for galaxies which fall
into both groups. Its noteworthy that the ``Both'' subset
is more massive and bluer, than the other two. This suggests
that the ``Both'' group is not just a random sample from
the other two groups. Interestingly, the overlap between
the two $x<0.1$ groups (that is, the ``Both'' group) is
only about 24 per cent, underlining that an morphology
heavily constrained by the SED does not imply a SED
heavily constrained by morphology, and vice versa.

We produce a contingency table for the mutually 
constrained group (MCG) but it does not reveal significant
over indexing and suggests instead a heterogeneous mix
of galaxies.
\begin{figure}
  \centering
  \includegraphics[width=8.6cm]{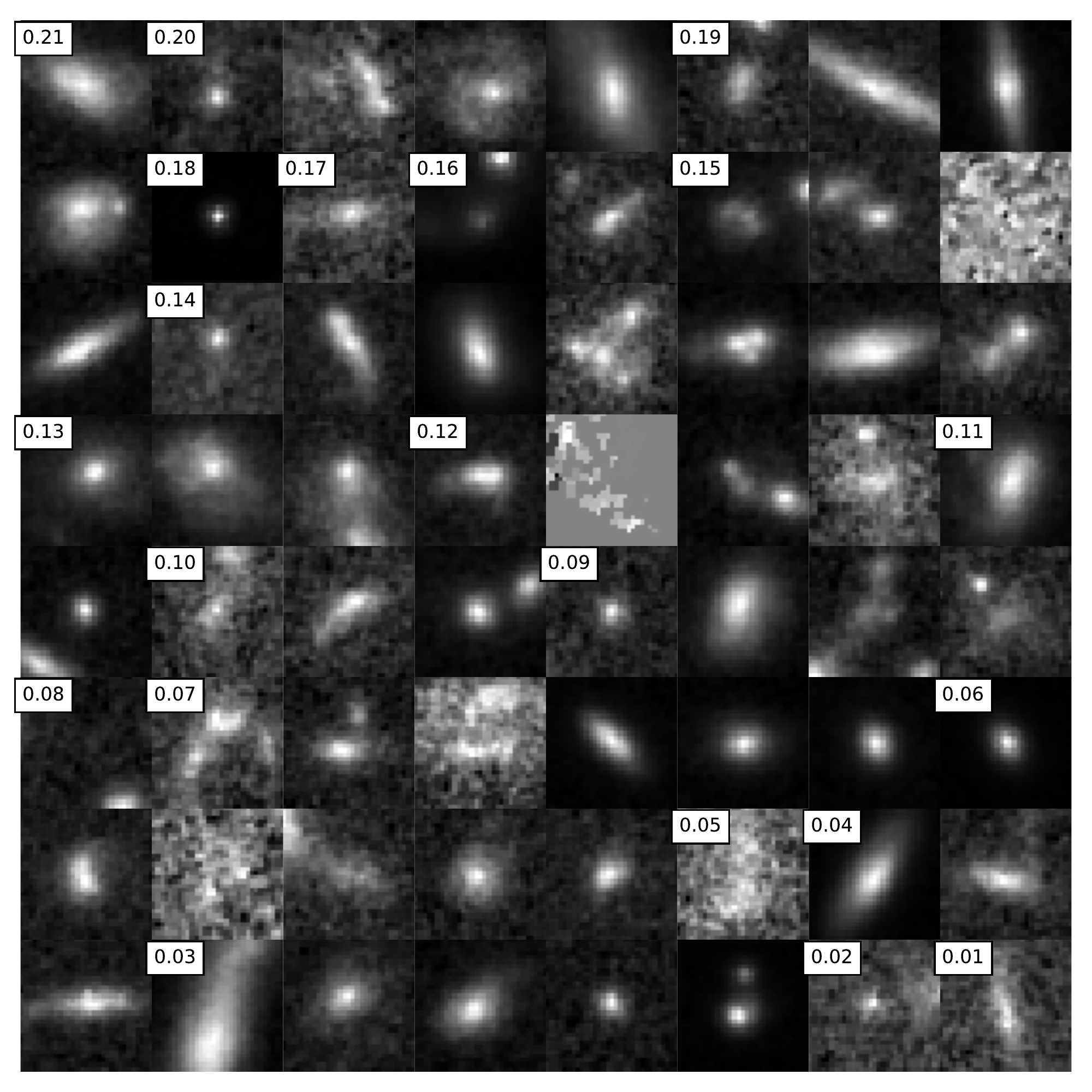}
  \caption {
    The partition exemplars ordered by effect size on a
    ridge classifier. The thumbnails are laid out left
    to right, top to bottom. The numbers indicate the
    difference in coverage when the feature associated
    to the exemplar is randomly permuted. Labels are
    omitted where they would be the same as the last
    label in the sequence. Objects with the greatest
    effect size typically fall into partitions containing
    fewer than 50 galaxies and appear more morphologically
    disturbed or clumpy compared with later exemplars.
  }
  \label{ridge-regression}
\end{figure}
We instead search for common visual features by training
a classifier to distinguish between members of the MCG and
the rest of the population. A novelty here is that we use
the distances to the partition exemplars as features thus
allowing feature importances to be interpreted in terms of
similarity of MCG members to particular exemplars. We use
a "ridge" classifier \citep[][see Appendix
\ref{ridge} for a short overview]{hoerl1970ridge} with 100-fold 
cross-validation (see Appendix \ref{kfold} for a short 
overview) to confirm an average classification accuracy\footnote{Since the MCG is small compared to the population,
balanced classification accuracy is used, which takes into
account the class frequencies to produce an accuracy figure
centered on 50 per cent.} of about 64 per cent, which
suggests that some visual features are more abundant in
the MCG than the general population.

To further investigate which exemplars most affect the
model, we randomly shuffle each feature in turn and
measure how much the intervention affects coverage (the
fraction of true positives correctly identified). This
procedure is known as permutation feature importance
\citep[][see Appendix \ref{importances}
for a short overview]{breiman1996bagging}. Bigger changes in coverage 
indicate that the feature is more important for picking
out MCG members. Figure \ref{ridge-regression} shows the
partition exemplars ordered by effect on coverage.
Based on these results, there are several features that
are particularly important in the MCG. Objects with the
greatest effect typically fall into partitions 
containing fewer than 50 galaxies and appear more 
morphologically disturbed or clumpy compared with later
exemplars. A plausible explanation is that the MCG objects
are affected by processes which have a simultaneous effect
on both the SED and morphological properties of the galaxy,
likely on short timescales, for example mergers, interactions
or clumpy star formation. This is particularly clear to see
when comparing the exemplars found in Figure \ref{group-comp}
to those found in Figure \ref{ridge-regression}. We leave
further analysis, classifier improvement and better
description of the MCG for future work.

\subsection{Implications for morphology selection by colour}

The classification of morphology using colour works by 
establishing constraints of the form $a < y < b$ for 
one or more colours which best separate the target
population from the rest. For example, \cite{strateva2001color}
use rest-frame $U-R \ge 2.22$ to separate early-type
galaxies from the rest with a claim of 80 per cent
coverage and 62 per cent purity (the overall fraction
correctly classified). The process of classifying by
colour can be iterative, with each subsequent threshold
attempting to strike a balance between purity (the rate
of true positives) and coverage. This bears direct analogy
to the operation of a decision tree classifier 
\citep[][see Appendix \ref{trees} for a short
overview]{breiman2017} which on each iteration -- in the binary case
-- picks the best of its features according to some
coverage/purity criterion to make a threshold cut, repeating
the process at every partition, until there are no more
gains to be made or some limiting condition -- such as
maximum depth -- has been reached. We can use the parallel
between selection by colour as practiced by astronomers
and the operations of a decision tree classifier, together
with the empirical partitions and SED space derived above,
to investigate whether we can establish an upper bound to
the efficacy of methods that use colour selection to
classify morphology.
\begin{table}
  \centering
\begin{tabular}{rlrrr}
\hline
   Part. & Best redshift range   &   \# &   Coverage &   Purity \\
\hline
       0 & 0.94-1.44 & 278 &       0.6  &       0.61 \\
       1 & 0.53-1.03 & 336 &       0.4  &       0.64 \\
       2 & 1.91-2.41 & 190 &       0.47 &       0.62 \\
       3 & 0.78-1.28 & 239 &       0.67 &       0.59 \\
       4 & 0.85-1.35 & 133 &       0.55 &       0.62 \\
       5 & 0.05-0.55 &  73 &       0.62 &       0.61 \\
       6 & 0.80-1.30 &  55 &       0.71 &       0.67 \\
       7 & 2.24-2.74 &  50 &       0.34 &       0.56 \\
       8 & 0.59-1.09 &  50 &       0.76 &       0.62 \\
       9 & 0.23-0.73 &  96 &       0.91 &       0.92 \\
      10 & 0.27-0.77 &  54 &       0.57 &       0.6  \\
\hline
\end{tabular}
  \caption {
   The best redshift range, sample size, coverage and
   purity achieved by a decision tree classifier based
   on colour features, for each partition. Purity is
   notably below 64 per cent for most partitions
   implying an upper bound limit to colour based
   methods.
  }
  \label{dec-tree}
\end{table}

For each empirical partition, we train a decision tree
classifier with a maximum depth of 6 using all the possible
colours in the $UBVRIJK$ bands as features: this equates to
looking for an optimal set of thresholds on up to six colours.
We repeat the fitting for every redshift window with a width
of 0.5, in which the partition has at least 50 members. That
is, for every partition, we try to find a redshift slice in
which some set of up to six colour constraints best separate
galaxies belonging to that partition from the rest. We use
10-fold cross-validation to establish the key metrics. The
table in Figure \ref{dec-tree} shows optimal redshift,
sample size, coverage and purity for all partitions big
enough to test. With the exception of Partition 9, the
purity of classifications is around the 60--70 per cent
level. Several partitions have coverage of at or above 75
per cent but it is 60 per cent or below for most partitions.
The 11 partitions tested span more than 75 per cent of the
data so it appears that morphological classification using
colour results in generally poor efficacy as measured by
coverage and purity.

\begin{table}
  \centering
\begin{tabular}{rlrrr}
\hline
   Part. & Best redshift range    &   \# &   Coverage &   Purity \\
\hline
       0 & 0.00-0.50 &  60 &       0.62 &       0.61 \\
       1 & 0.58-1.08 & 327 &       0.54 &       0.64 \\
       2 & 0.16-0.66 &  73 &       0.82 &       0.73 \\
       3 & 0.05-0.55 & 109 &       0.7  &       0.62 \\
       4 & 0.87-1.37 & 134 &       0.68 &       0.68 \\
       5 & 0.79-1.29 &  92 &       0.74 &       0.71 \\
       6 & 0.71-1.21 &  63 &       0.76 &       0.73 \\
       7 & 1.89-2.39 &  50 &       0.46 &       0.53 \\
       8 & 0.57-1.07 &  50 &       0.8  &       0.68 \\
       9 & 0.03-0.53 &  78 &       0.98 &       0.9  \\
      10 & 0.33-0.83 &  50 &       0.66 &       0.65 \\
\hline
\end{tabular}
  \caption {
    The best redshift range, sample size, coverage and
    purity achieved by a ridge classifier on rest-frame
    band features, for each partition. Both coverage
    and purity are improved for almost all partitions
    when compared to the decision tree classifier on
    colour features.
  }
  \label{col-regression}
\end{table}

Efficacy can be improved by abandoning colours in favour
of any linear combination of rest-frame bands. We fit
a Ridge classifier using the SEDs as features and repeat
the fitting for every redshift window with a width of 0.5
in which the partition has at least 50 members. That is, 
for every partition, we try to find a redshift slice in which 
some combination of bands best separates galaxies belonging
to that partition from the rest. As before, we use 10-fold
cross-validation to establish the key metrics. Table 
\ref{col-regression} shows the optimal redshift range, sample
size, coverage and purity for all partitions tested. Its
noteworthy that almost all coverage and purity figures are
improved by using all the bands, tipping purity for many
partitions past the 70 per cent level.

Since the empirical partitions are homogeneous morphologies
which together span the whole dataset, one would expect that
coarser groupings (e.g. early versus late types) could be
produced by pooling partitions together into fewer
morphological classes. However, since the efficacy -- as
measured by coverage and purity -- is relatively low for
both methods, it is unlikely that pooling would result in
better morphological classification, and could make it worse
by decreasing in-class homogeneity. It should be further
noted that the numbers for coverage and purity presented
here are effectively \textit{upper bounds}, since we report
only the results for the optimal redshift slice. Hence, 
significantly worse efficacy could be expected in other
redshift ranges. In general, therefore, pure morphological classes cannot be selected very effectively using colours. 

\section{Summary}\label{conclusion}

In this work we have introduced an empirical methodology for 
the analysis of how galaxy SEDs and morphologies constrain
each other, using the vector space building techniques for 
galaxy surveys laid out in \cite{uzeirbegovic2020}. Our main
results can be summarised as follows:

\begin{enumerate}
    \item Two galaxies with very similar SEDs are around 3 times more likely to also be most morphologically similar, compared to the null hypothesis that SED and morphology are independent. Massive red ellipticals are especially likely to be well-identified their SEDs.
    
    \item Two morphologically similar galaxies are slightly under 3 times more likely to be most SED similar, compared to the null hypothesis that SED and morphology are independent. However, morphology constrains the SED more strongly than vice versa on average (with around a 7 per cent improvement in the space fraction).
    
    \item Disturbed or interacting systems -- systems which are experiencing processes like mergers that affect both the SED and the morphology simultaneously -- are prominent amongst galaxies for which the SED and morphology are mutually constraining (i.e. in which the SED implies a relatively similar morphology and vice versa).
    
    \item No combination of colour cuts is able to strongly constrain galaxy morphology. On average, purity is around 64 per cent, if up to 3 colours are used to try and select homogeneous morphological classes. While the results can be improved by considering linear combinations of the whole SED, the improvements are not significant and purity levels for most morphological classes remain at the 70 per cent or lower.

\end{enumerate}

\section*{Data availability}

\noindent Code for all results and figures can be found at
\url{https://emiruz.com/similarity}.

\section*{Acknowledgements}
SK acknowledges support from the STFC [ST/S00615X/1] and a
Senior Research Fellowship from Worcester College Oxford.
This work is based on observations taken by the CANDELS
Multi-Cycle Treasury Program with the NASA/ESA HST, which
is operated by the Association of Universities for Research
in Astronomy, Inc., under NASA contract NAS5-26555. This
research made use of Astropy, 
\footnote{http://www.astropy.org} a community-developed
core Python package for Astronomy \citep{astropy:2013, astropy:2018}.

\section{Appendix}

\subsection{Ridge classifier} \label{ridge}

A ridge regression \citep{hoerl1970ridge} is an alternative
to ordinary least squares regression for fitting multiple
regression models which is particularly suitable when the
independent variables are highly correlated. The regression
is made into a binary classifier by coding the dependent
variable as either 1 (true) or -1 (false) and then conducting
a regression. Ridge regression is well suited to the task at
hand because (1) it out-performs more complex algorithms
which we have also tried, and (2) we are dealing with a
large number of correlated independent variables.

\subsection{Permutation feature importance} \label{importances}

The permutation feature importance \citep{breiman1996bagging}
is the decrease in a metric used to score a model (coverage in
our case) when a single independent variable is randomly permuted.
The permutation breaks the relationship between the independent
and the dependent variables, allowing the drop in the scoring
metric to be used as an indicator of the extent to which the
model depends on the feature. Since permutations are random,
the processes must be repeated many times per independent 
variable to calculate an average effect.

\subsection{Decision trees} \label{trees}

Classification and regression trees \citep{breiman2017} are an 
umbrella term for various ways to construct decision trees for
the purposes of classification and regression. We utilise a
relatively basic procedure for our binary classification case,
wherein independent variables are evaluated one at a time,
and the one which maximises a metric such as information gain
(difference in information entropy) is selected as a splitting
criteria. For each split the procedure is recursively repeated
until some stopping criteria -- such as maximum tree depth or
too small a gain -- is reached.

\subsection{$K$-fold cross validation} \label{kfold}

$K$-fold cross validation refers to the splitting of the data
into $K$ pieces wherein each piece is used for testing in turn,
whilst the $K-1$ other pieces are used for training. The 
individual evaluations may be combined together to produce an
overall result.

\subsection{Affinity propagation}\label{affinity}

Affinity propagation \citep{dueck2009affinity} is a clustering
algorithm which partitions data around a set of "exemplars" by
solving the optimisation problem presented in Equation \ref{apeq}.
The problem is known to be intractable, but affinity propagation
uses a way of reformulating the problem known as "message passing",
such that a solution can be approximated by iteration. More
particularly, affinity propagation makes extensive use of the
sum-product rule \citep{pearl1982reverend} to minimise the number
of computations required on each iteration in order to make the
algorithm tractable. Key benefits of affinity propagation for
our problem are that (1) it is exemplar based, and (2) that
it discovers the number of exemplars.

\bibliographystyle{mnras}
\bibliography{paper}

\end{document}